\documentclass[a4paper,twocolumn,11pt]{quantumarticle}
\pdfoutput=1
\usepackage[dvipsnames]{xcolor}

\usepackage[utf8]{inputenc}
\usepackage[english]{babel}
\usepackage[T1]{fontenc}
\usepackage{dsfont}

\usepackage{amssymb}
\usepackage{amsmath}
\usepackage{braket}
\usepackage{hyperref}
\usepackage{listings}
\usepackage{tikz}
\usepackage{lipsum}
\definecolor{codegreen}{rgb}{0,0.6,0}
\definecolor{codegray}{rgb}{0.5,0.5,0.5}
\definecolor{codepurple}{rgb}{0.58,0,0.82}
\definecolor{backcolour}{rgb}{0.95,0.95,0.92}

\lstdefinestyle{mystyle}{
    backgroundcolor=\color{backcolour},
    commentstyle=\color{codegreen},
    keywordstyle=\color{magenta},
    numberstyle=\tiny\color{blue},
    stringstyle=\color{codepurple},
    basicstyle=\ttfamily\footnotesize,
    breakatwhitespace=false,
    breaklines=true,
    captionpos=b,
    keepspaces=true,
    numbers=left,
    numbersep=5pt,
    showspaces=false,
    showstringspaces=false,
    showtabs=false,
    tabsize=2
}

\usepackage[normalem]{ulem} 

\makeatletter

\begin{document}
\title{The SpinPulse library for transpilation and noise-accurate
simulation of spin qubit quantum computers}

\author{Beno\^it Vermersch}
\affiliation{Quobly, 38000 Grenoble, France}
\affiliation{Universit\'e Grenoble Alpes, CNRS, LPMMC, 38000 Grenoble, France}
\orcid{https://orcid.org/0000-0001-6781-2079}
\author{Oscar Gravier}
\affiliation{Quobly, 38000 Grenoble, France}
\affiliation{Universit\'e Grenoble Alpes, CEA-L\'eti, F-38054 Grenoble, France}
\author{Nathan Miscopein}
\affiliation{Quobly, 38000 Grenoble, France}
\author{Julia Guignon}
\affiliation{Quobly, 38000 Grenoble, France}
\affiliation{Center for Quantum Science and Engineering, Ecole Polytechnique Fédérale de Lausanne (EPFL), CH-1015 Lausanne, Switzerland}
\author{Carlos Ramos Marim\'on}
\affiliation{Quobly, 38000 Grenoble, France}
\author{Jonathan Durandau}
\affiliation{Quobly, 38000 Grenoble, France}
\author{Matthieu Dartiailh}
\affiliation{Quobly, 38000 Grenoble, France}
\author{Tristan Meunier}
\affiliation{Quobly, 38000 Grenoble, France}
\author{Valentin Savin}
\affiliation{Quobly, 38000 Grenoble, France}
\affiliation{Universit\'e Grenoble Alpes, CEA-L\'eti, F-38054 Grenoble, France}

    \begin{abstract}
       We introduce \texttt{SpinPulse}, an open-source \texttt{python} package 
       for simulating spin qubit-based quantum computers at the pulse-level. 
       \texttt{SpinPulse} models the specific physics of spin qubits,
        particularly through the inclusion of classical non-Markovian noise.
         This enables realistic simulations of native gates and quantum circuits,
          in order to support hardware development.
           In  \texttt{SpinPulse}, a quantum circuit is first transpiled into the native gate set of our model and then converted to a pulse sequence. This pulse sequence is subsequently integrated numerically in the presence of a simulated noisy experimental environment.
We showcase workflows including transpilation, pulse-level compilation, hardware benchmarking, quantum error mitigation, and large-scale simulations via integration with the tensor-network library \texttt{quimb}. We expect \texttt{SpinPulse} 
to be a valuable open-source tool for the quantum 
computing community, fostering efforts to devise 
high-fidelity quantum circuits and improved 
strategies for quantum error mitigation and correction.
    \end{abstract}

    \section{Introduction}

    The power of quantum computing can only be envisioned
    when the effect of noise
    is understood and reduced to a tolerance threshold.
    This statement applies
    obviously to the long-standing goal of fault-tolerant quantum computing
    where logical errors can be set to an arbitrarily small level using quantum
    error correction~\cite{fowler_surface_2012}, but also to quantum error mitigation in the context of quantum simulation~\cite{cai_quantum_2023}.
       Classical simulations of quantum circuits allow us to assess the performance of quantum computing in presence of noise in certain parameter regimes, and to develop noise-specific quantum error correction/mitigation strategies.
  \texttt{SpinPulse} is an open-source python package that addresses these questions in the context of spin qubits~\cite{burkard_semiconductor_2023}.

The description of noise takes different flavors depending
    on the type of hardware under study~\cite{preskill_lecture_2018}.
    Under Markovian noise, a quantum system evolves via a Lindblad master
    equation, which can be integrated to yield a quantum channel defined via
    Kraus operators.
    These operators specify the noise and can be incorporated easily
    in any simulator tool, such as \texttt{qiskit\_aer}~\cite{javadi_qiskit_2024}.
 Simulations can be performed either at the density matrix level or using
    the equivalent representation provided by stochastic quantum
     trajectories~\cite{preskill_lecture_2018}.

    In the context of spin qubits,
     the non-Markovian character of the noise
    makes the simulation more challenging, as a description with fixed Kraus
    operators is no longer possible~\cite{burkard_semiconductor_2023}.
    This does not mean the noise is more difficult
    to understand and mitigate.
    For instance, the presence
    of time correlated terms in the system Hamiltonian,
    at the heart of the non-Markovian character of the dynamics,
    can be cancelled out efficiently during transpilation using dynamical
     decoupling strategies~\cite{lidar_review_2014}.

     In this context, our package \texttt{SpinPulse} provides a complete workflow for simulating and optimizing noisy spin qubit quantum circuits.
     Through sequential transpilation modules,
     we provide a pulse-level description of a circuit,
     i.e., a space-time representation of a realistic experimental sequence,
     to which we can attach
     random realizations of a non-Markovian noise model.
     These circuits can then be converted to a gate-level description,
     and simulated with any gate-level simulator.
    This allows us to predict essential quantum circuit metrics, such as average
    gate fidelities, averaged correlation functions, etc., that we will
    illustrate throughout this paper.
    We will show in particular how the package proves useful in
   quantifying the effect of noise in terms of quantum channels, and how
    to incorporate dynamical
    decoupling strategies efficiently.
    This article is aimed to be self-contained: we first address the physical details of a
    simplified spin qubit quantum computing model.
    We then describe our construction of the package implementing this model.

  Note finally that pulse-level simulators have been made available in other contexts.
      In the neutral-atom case, pulse-level simulators are ubiquitous as the device
      is often operated in analog mode~\cite{silverio_pulser_2022,quera_bloqade_2023}.
      Unfortunately, these tools are not related to a spin qubit
      native set, as described in our model Sec.~\ref{sec:model}.
      In the superconducting qubit context, \texttt{qiskit\_dynamics}~\cite{puzzuoli_qiskit_2023}
      offer parametrizable pulse-level descriptions, but do not include
      non-Markovian noise models.
      By providing noise-accurate simulation of spin qubit quantum computers,
      we believe that our package is a valuable contribution to the
      community. 
      In our conclusion, we describe how contributions
      can be modularly implemented to pursue this effort.

  \section{General overview of the package}\label{sec:organization}

\begin{figure*}
  \begin{center}
  \includegraphics[width=0.9\textwidth]{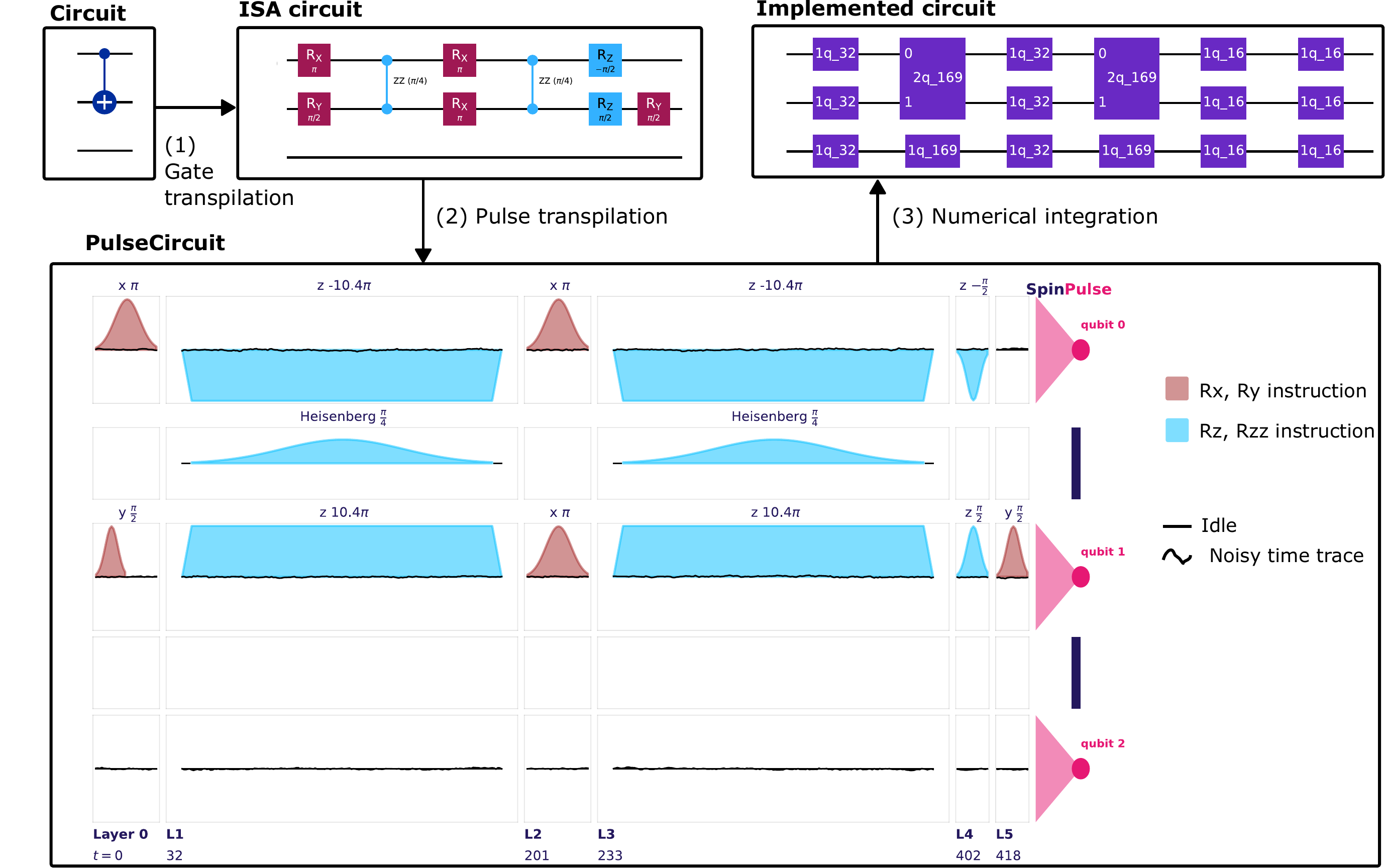}

  \end{center}
   \caption{Organization of our package with the three steps (1),
   (2), (3) described in Sec.~\ref{sec:organization}. A \texttt{qiskit}
   circuit is converted to an Instruction Set Architecture (ISA) circuit using an instance of
   \texttt{HardwareSpecs} class, which is equipped with a hardware model
   specific
   \texttt{qiskit} transpiler. This ISA circuit is then converted to
   a \texttt{PulseCircuit} instance that describes a pulse-level
   circuit in the presence of additional noise terms. This circuit can then be numerically
   integrated to form a noise-accurate \texttt{qiskit} circuit.}
   \label{fig:package}
   \end{figure*}

The structure of our package is schematically shown in Fig.~\ref{fig:package}.
Our overall goal is to mimic the execution of a quantum circuit. This happens in
three steps that we will describe in detail in Sec.~\ref{sec:description}.

\textbf{
  (1) Gate transpilation---}
  First, a quantum circuit written in \texttt{qiskit}~\cite{javadi_qiskit_2024} language is transpiled into the
native gate set of our model, described in Sec.~\ref{sec:model}.

\textbf{
  (2) Pulse transpilation---}
Based on our model and corresponding hardware specifications, the package
expresses each gate into a pulse sequence describing a time-dependent
 Hamiltonian. One can attach at this stage stochastic noisy contributions
 to the Hamiltonian according to our noise model. The resulting object is defined as a
 \texttt{PulseCircuit}, which is the central structure type of our package.

 \textbf{(3) Gate integration---}
Calculating the evolution operator associated with each pulse sequence, we obtain the noisy
version of each gate of the original circuit. This forms a noisy quantum circuit
that we can simulate with \texttt{qiskit\_aer} or any other gate-based simulation tool,
such as \texttt{quimb}~\cite{gray_quimb_2018}.
The average evolution of such a circuit over many realizations of the noise model attached during (2) pulse
transpilation corresponds to the quantum channel associated with our noise model.

To explain how the package implements the aforementioned steps (1-3), we first
need to describe our model.

    \section{A quantum computing model for spin qubits}\label{sec:model}

    Our gate and noise model is inspired by the one presented in
    Ref.~\cite{gravier_simulated_2025}, and corresponds
    to a generic formulation of spin qubits trapped in quantum
    dots, see also Ref.~\cite{burkard_semiconductor_2023}.
  The present formulation aims to be at the same time powerful and simple,
  in the sense that it provides a universal quantum gate set based on analytical
  relations between gate parameters and Hamiltonian pulse specification. This will be useful to define
  our chain of transpilation in the next section.

  \subsection{Hamiltonian model}
    We consider a set of $i=0,\dots,N-1$ qubits of Larmor frequencies $\omega$. 
    Each of them is subject to a Hamiltonian 
\begin{align}
\label{eq:hi}
h_i(t)&=\frac{B_i(t)}{2}\Big (\cos\big(\varphi_i(t)\big)X_i+\sin\big(\varphi_i(t)\big)Y_i\Big)
\nonumber \\
&+\frac{\delta \omega_i(t)}{2}Z_i\, ,
\end{align}
in the frame rotating with $\omega$, and where the complex driving field in the static frame reads
\begin{equation}
  \mathbf{B_i}(t)=B_i(t)\exp \big(-i \omega t+i\varphi_i(t)\big)\, ,
\end{equation}

\noindent and $\delta \omega_i(t)$ is a DC-generated Stark shift.

The interaction between qubits are expressed as  $H_2(t)=\sum_{<i,j>}h_{i,j}(t)$ with the Heisenberg Hamiltonian (both in the static and rotating frame)
\begin{equation}
  \label{eq:heisenberg}
  h_{i,j}(t)=\frac{J_{i,j}(t)}{2}(X_iX_j+Y_iY_j+Z_iZ_j)\, .
\end{equation}
In addition to this noiseless Hamiltonian,
we will consider noisy Hamiltonian terms in Sec.~\ref{sec:noise_model}.

\subsection{Units and time discretization}

In what follows, all energies and coupling constants are written
in units of $\hbar$, i.e., they can be interpreted as angular frequencies.
As our model is attached to a computer code, we also find
convenient to describe dynamics with a fixed sampling time $t_s$,
which can be considered as our unit of time.
This means we express the time $t$ as a dimensionless vector $t=1,2,\dots,t_{\max}$ in units of $t_s$.
The time $t_s$ thus corresponds to
the time resolution of our simulated numerical experiments.
For instance, if $t_s$ corresponds to $1$ ns, we will express angular frequencies in
rad.(ns)$^{-1}$ and will be able to resolve all dynamics corresponding to frequencies below the GHz scale.

\subsection{Quantum gates}
Single- and two-qubit gates are implemented by evolving the system according
to the above Hamiltonians, using suitably parametrized pulses $B_i(t)$,
$J_{i,j}(t)$.

\subsubsection{Single-qubit gates}

Our single-qubit gate set is composed of the three parametrized gates $R_{\sigma}(\theta)=\exp\big(-i\frac{\theta}{2} \sigma\big)$, where $\sigma=X,Y,Z$.

Setting $\delta \omega_i(t)=0$ and $\varphi_i(t)=0$
in Eq.~\eqref{eq:hi}, evolving the qubit $i$ in time with $h_i(t)$
gives rise to the $R_X$ gate
\begin{equation}
R_X\big(\theta_i(t)\big)=\mathcal{T}\exp\left(-i\int h_i(t)\right),
\end{equation}

\noindent with $\theta_i(t)=\int_0^t B_i(t')dt'$.
Remember that the integral is considered to be taken
on a discretized function of the dimensionless time vector $t$, i.e,  $\int_0^t(\cdot) dt=\sum_{t'=1}^t(\cdot). $
Similarly, we obtain an $R_Y(\theta_i)
$ gate using $\varphi_i(t)=\pi/2$.
For the $R_Z(\theta_i)$ gate we set instead
$B_i(t)=0$, and have $\theta_i=\int_0^t \delta \omega_i(t')dt'$.

\subsubsection{Two-qubit gate}\label{sec:twoq_gate}
 Our two-qubit gate is the parametrized 
\begin{equation}
R_{ZZ}(\theta)=\exp\big(-i\frac{\theta}{2} ZZ\big).
\end{equation}
For a qubit pair $(i,j)$, it can be implemented by considering the time-dependent Hamiltonian
\begin{align}
 h(t) &
 = \frac{J_{i,j}(t)}{2}(X_iX_j+Y_iY_j+Z_iZ_j)
 \nonumber \\
 &+\frac{\delta \omega_i(t)}{2}Z_i
 +\frac{\delta \omega_j(t)}{2}Z_j\, .
\end{align}
 In the following, we drop the $i,j$ indices when possible.
 This can be put in a form suitable for a Schrieffer-Wolff (SW) 
 transformation~\cite{bravyi_schriefferwolff_2011}:
 \begin{align}
     h(t)&=h_0(t)+V(t), \\
     h_0(t)&=\frac{J(t)}{2}Z_iZ_j+\frac{\delta \omega_i(t)}{2}Z_i
 +\frac{\delta \omega_j(t)}{2}Z_j, \\
 V(t)&=\frac{J(t)}{2}(X_iX_j+Y_iY_j).
 \end{align}

The SW transformation is based on perturbation theory 
 and approximates the dynamics by an effective Hamiltonian
  $h_\mathrm{SW}(t)$ in the regime $J(t)\ll\Delta(t)$, with $\Delta(t)=\delta \omega_i(t)-\delta \omega_j(t)$.
 The procedure consists in defining a generator $S(t)$ of
  this transformation that satisfies
\begin{equation}
 V(t)+[S(t),h_0(t)]=0,
\end{equation}

 \noindent leading to 
 \begin{equation}
 S(t)=\frac{iJ(t)}{2\Delta(t)}(Y_iX_j-X_iY_j).
 \end{equation}
 
 The SW Hamiltonian is equal to
\begin{equation}
 h_\mathrm{SW}(t)=e^{S(t)}\Big(h_0(t)+\frac{1}{2}[S(t),V(t)]\Big)e^{-S(t)},
\end{equation}

 \noindent with the Stark-Shift terms
 \begin{equation}
h_\mathrm{sh}=[S(t),V(t)]=\frac{J^2(t)}{2\Delta(t)}(Z_i-Z_j)\, .
 \end{equation}

 The SW transformation gives an analytical recipe to realize an $R_{ZZ}$ gate.
 First, we ramp $\Delta(t)$ up from $0$ to a plateau value
 $\Delta$, while keeping $J(t)=0$.
 Then, we slowly drive the value of $J(t)$ from $0$ through a maximum value, and then return to $0$.
 According to the quantum adiabatic theorem, each wave-function amplitude in the $Z$ basis, which is the eigenbasis of the Hamiltonian at $t=0$, will rotate with the energy of the corresponding instantaneous eigenstate at later times (see App.~\ref{app:adiabatic} for a step-by-step derivation).
 This means that we obtain an evolution operator
\begin{equation}
 u(t)=\mathcal{T}\exp\left(-i\int_0^t h_0(t')+\frac{1}{2}[S(t'),V(t')]dt'\right) .
\end{equation}

The Stark-shift terms commute with the entangling term generated by $h_0(t')$, and we can write
 \begin{align}
  u(t)&=\mathcal{T}\exp\left(-i\int_0^t h_0(t')dt'\right)
  u_\mathrm{sh}(t)
  \nonumber \\
  &=R_{ZZ}\big(\theta(t)\big)
  v(t),
 \end{align}
 
 \noindent with angle $\theta(t) = \int_0^{t} J(t') dt'$.
 Here, $u_\mathrm{sh}(t)$ describes the evolution with the Stark-Shift terms
 $h_\mathrm{sh}(t)$, and $v(t)$ adds the remaining rotation
 induced by $\frac{\delta \omega_i(t)}{2}Z_i+\frac{\delta \omega_j(t)}{2}Z_j$.
 The effect of $v(t)$ can be removed exactly by considering a spin echo version
 of this gate~\cite{gravier_simulated_2025}, where we apply two $X$ pulses, i.e.,
\begin{eqnarray}
  \label{eq:rzz}
  (u(t/2)X_iX_j)^2=R_{ZZ}\big(\theta(t)\big)\, ,
\end{eqnarray}

\noindent where we used that $X_iX_jvX_iX_j=v^\dag$.

 \subsection{Noise models and dissipative evolutions}\label{sec:noise_model}

We model noise in the first approximation by classical phase noise~\cite{burkard_semiconductor_2023}, which affects the qubits by an additional contribution
$\delta h_i(t) = \frac{\epsilon_i(t)}{2} Z_i$ to
the single-qubit Hamiltonian $h_i(t)$ [Eq.~\eqref{eq:hi}].
Here, classical means that the generators of the noise term are
classical random fields $\epsilon_i(t)$. This allows us to access
the average evolution of our circuit by performing
stochastic averages over random realizations of $\epsilon_i(t)$.
To be specific, we define the average quantum channel
of a quantum circuit~\cite{preskill_lecture_2018}
\begin{equation}
 \mathcal{E}(\rho)
 =
\mathbb{E}\left[ u\rho u^\dag\right],
\end{equation}
where here $u$ is the generator of the total Hamiltonian including
noise, and $\rho$ the density matrix of the system (or subsystem) of
the quantum computer.

It will also be convenient for what follows to  use the superoperator representation
$\mathcal{S}=\mathbb{E}\left(u^*\otimes u\right)$ of the channel $\mathcal{E}$, and the $\chi$ matrix
defined through the relation~\cite{wood_tensor_2015}
\begin{equation}
  \mathcal{E}(\rho)
  =\sum_{i,j}\chi_{i,j}P_i \rho P_j\, ,
 \end{equation}
 where $\{P_i\}$ are strings of Pauli matrices.

An important figure of merit is the average fidelity $\mathcal{F}$ (with respect to Haar random initial state)
with respect to a reference circuit \mbox{$\mathcal{S}_0=u_0^*\otimes u_0$}, given by~\cite{wood_tensor_2015}
\begin{equation}
  \label{eq:gate_fidelity}
  \mathcal{F}
  = \frac{d\mathcal{F}_\mathrm{pro}+1}{d+1},
\end{equation}
with $d$ the dimension of $u$, and the process fidelity
\begin{align}
\mathcal{F}_\mathrm{pro}
&=
\frac{1}{d^2}\mathrm{Tr}
\left(
\mathcal{S}_0^\dag
\mathcal{S}
\right)=
\frac{1}{d^2}\mathbb{E}
\left[
|\mathrm{Tr}(u_0^\dag u)|^2
\right]\,.
\end{align}

\subsubsection{Channels under idle noise, and Ramsey experiments}\label{sec:idle_noise}
Let us consider a single-qubit, dropping the index $i$ for simplicity, and study the evolution under
 $\delta h(t)$
with $u=e^{-i \frac{\phi(t)}{2}Z}$, and $\phi(t)=\sum_{t'\le t}\epsilon(t')$.
The average channel reads
\begin{eqnarray}
  \label{eq:def_channel}
  \mathcal{E}(\rho) = \mathbb{E}[u\rho u^\dag]\, .
\end{eqnarray}
Expanding the expression for $u\rho u^\dag$ yields
\begin{align}
  \label{eq:chan_ref}
  u\rho u^\dag
    &= \frac{1+\cos\big(\phi(t)\big)}{2}\rho
    + i\frac{\sin\big(\phi(t)\big)}{2}
    (\mathds{1}\rho Z - Z\rho\mathds{1})
   \nonumber \\
   & + \frac{1-\cos\big(\phi(t)\big)}{2}Z\rho Z\,.
\end{align}
The average channel thus reads
\begin{eqnarray}
  \label{eq:chan_average}
    \mathcal{E} (\rho)
    = \frac{1+C(t)}{2}\rho
    +\frac{1-C(t)}{2}Z\rho Z
\end{eqnarray}
where we assumed a symmetric noise such that $\mathbb{E}[\sin(\phi(t))]=0$, and defined
\mbox{$C(t)=\mathbb{E}[\cos(\phi(t))]$}.

The quantity $C(t)$ can be interpreted as the Ramsey contrast.
 Indeed, for an initial state, $\rho=\ket{+}\bra{+}$, $\ket{+}=H\ket{0}$,
we would measure
\begin{equation}
\braket{X} = \mathrm{Tr}\big(\mathcal{E}(\rho)X\big)=\frac{1+C(t)-(1-C(t))}{2}=C(t).
\end{equation}
Noticing that
\begin{equation}
\mathcal{S}=\frac{1+C(t)}{2}(\mathds{1}\otimes \mathds{1})
+
\frac{1-C(t)}{2}(Z\otimes Z)\, ,
\end{equation}
we also obtain the average fidelity
\mbox{$\mathcal{F}=\big(2+C(t)\big)/3$}.

\subsubsection{Noise types}\label{sec:noise_types}

Three types of noise are implemented: quasi-static, white, and pink.
\clearpage

\textbf{Quasi-static noise---}
In the case of quasi-static noise, we have a time-independent signal $\epsilon(t)=\epsilon$ where $\epsilon$ is chosen from a random normal distribution of standard deviation $\sigma$.
Thus,
$\phi(t) = \epsilon t$ and
\begin{equation}
C(t)=\mathbb{E}[\cos(\epsilon t)]=\exp(-t^2\sigma^2/2)\, ,
\end{equation}
which allows to identify a coherence time $T_2^*=\sqrt{2}/\sigma$.
As explained below, in our numerical simulations,
 the signal will be kept constant
only on finite time intervals
 of length $\texttt{segment\_duration}$, which
  should be taken equal or larger than the circuit duration. 

\textbf{White noise---}
In the case of white noise, we have a time-dependent signal where $\epsilon(t)$ is chosen from a random normal distribution of standard deviation $\sigma$.
This leads to a Ramsey contrast
\begin{align}
C(t)&=\mathbb{E}[\cos(\phi(t))]
=\mathbb{E}[\exp(-i\phi(t)/2)]
\nonumber \\
&=\prod_{t'=1}^t \mathbb{E}\big[\exp\big(-i \epsilon(t')\big)\big]
= \exp(-t\sigma^2/2)\, ,
\end{align}
which gives an exponential decay associated with a coherence time $T_2^*=2/\sigma^2$.

\textbf{Pink noise---}
We consider time traces of the form
$\epsilon(t)=2\pi \sqrt{S_0} g(t)$, where $S_0$ is the spectral intensity
 (which has the dimension of a squared frequency,
 and is thus written in units of $t_s^{-2}$), and $g(t)$
is a normalized time trace with power spectral
 density $S_g(f)=1/f$, considering $f$ as the discretized frequency
 associated with the time vector $t$ introduced above.
Following Refs.~\cite{gravier_simulated_2025, yoneda_999-fidelity_2018}, we obtain
\begin{equation}
  C(t)\approx \exp\left(-t^2/(T_2^*(t))^2\right),
\end{equation}
with a \emph{time-dependent} value of the coherence time
\begin{equation}
\label{eq:T2vst}
T_2^*(t) = \frac{1}{2\pi\sqrt{S_0\log(\frac{1}{f_{\min}t})}},
\end{equation}
where $f_{\min}$ is a low-frequency cutoff.
In the short-time limit $f_{\min}t\ll 1$, we can approximate
$T_2^*(t)\approx T_2^*(1)$ as an effective constant coherence time, 
and $C(t)$ as being approximately described by a Gaussian decay.

\subsubsection{Noisy gates under Larmor frequency noise}

Let us now analyze the effect of noise during
 certain rotation gates.

\textbf{$X$ gate}---
As a first example, we consider the Hamiltonian
\begin{equation}
  h_i=\frac{B_0}{2}X+\frac{\epsilon}{2}Z\, ,
\end{equation}
that implements a $X$ gate with a square pulse of amplitude $B_0$ during
the gate time $t=\pi/B_0$, in the presence of quasi-static noise, i.e.,
$\mathbb{E}(\epsilon)=0$ and $\mathbb{E}(\epsilon^2)=\sigma^2$. 
Following the derivation given in App.~\ref{app:noise_channels}, we obtain
\begin{align}
    \mathcal{E}_{X}(\rho)
    &=
    \left(
    1-\frac{\sigma^2}{B_0^2}
    \right)
    X \rho X
    +\frac{\pi \sigma^2}{4B_0^2} (I\rho X+X\rho I)
    \nonumber \\
    &+
    \frac{\sigma^2}{B_0^2}Z\rho Z.
\end{align}
The first and third terms are Pauli channels corresponding respectively
 to the ideal $X$ gate,
  and to an unwanted phase flip occurring with probability $\sigma^2/B_0^2$. The second term corresponds to a non-Pauli process.

For white and pink noise, we will see in Sec.~\ref{sec:advanced}
that the quantum channel can differ significantly.

\textbf{$R_{ZZ}$ gate}---
We now consider the noisy $R_{ZZ}$ gate.
Under the conditions of the SW transformation $J(t)
\ll \Delta (t)$ described above, and assuming that the $X$ gates
 used in Eq.~\eqref{eq:rzz} are not affected by the noise, the effect of the noisy terms
$\epsilon_1(t)$, $\epsilon_2(t)$ can be calculated analytically 
(regardless of the type of noise: quasi-static, white, pink).
The derivation presented in App.~\ref{app:noise_channels},
ends up with the expression of the quantum channel
 \begin{equation}
  \mathcal{E}_{ZZ}(\rho)
  =
  R_{ZZ}(\theta)
\big(
  \Phi\otimes\Phi(\rho)
\big)
R_{ZZ}^\dag(\theta)
 \end{equation}
with 
 \begin{align}
    \Phi (\sigma)
    &= \frac{1+C_\mathrm{se}(t)}{2}\sigma
    +\frac{1-C_\mathrm{se}(t)}{2}Z\sigma Z,\end{align}
and spin-echo contrast
\begin{align}
C_\mathrm{se}(t)&=\mathbb{E}
\left[\exp
 \left(-i\phi_{k}(t) \right)
  \right]
  \nonumber \\
  \phi_{k}(t)&=
 \int_0^{t/2}
\epsilon_{k}(t')dt'
  -
  \int_{t/2}^t\epsilon_{k}(t') dt'.
 \end{align}
Hence, the noise-channel is the application of the spin-echo noise ``idle term''
$\Phi\otimes\Phi$ followed by the ideal noiseless gate evolution.
This result is a consequence of the noisy Hamiltonian terms commuting with the gate generator
$Z_iZ_j$.
This derivation shows in particular that the $R_{ZZ}$ gate is robust to quasi-static noise,
as it is completely suppressed via spin-echo.

\subsubsection{Application to dynamical decoupling}
Dynamical Decoupling (DD) aims at applying a sequence of pulses on a qubit that is idle during
a given time $t$, in order to suppress noisy contributions~\cite{lidar_review_2014}.
This technique is particularly useful for non-Markovian noise, where time correlations
correspond to low frequency spectral components~\cite{cywinski_how_2008}.
DD sequences can be interpreted as a filter removing such low frequencies from the noise signal.

\textbf{Dynamical decoupling with instantaneous pulses---}
Let us first consider the possibility of applying instantaneous
noiseless pulses during the idle sequence.
The simplest DD sequence is the spin echo
\begin{equation}
 u_\mathrm{se}(t)= Xu\left(\frac{t}{2},t\right) Xu\left(0,\frac{t}{2}\right),
\end{equation}
which we also saw in the context of the $R_{ZZ}$ gate implementation described
previously.
Here $u(t_1,t_2)$ denotes idle evolution from time $t_1$ to $t_2$. 
We obtain exact noise cancellation  $u_\mathrm{se}(t)=\mathds{1}$ in the case of
 quasi-static noise, where
 \mbox{$u\left(0,\frac{t}{2}\right)=u\left(\frac{t}{2},t\right)=\exp\left(-i\frac{\epsilon t}{4}Z\right)$}.

With non-stationary noise,
noise cancellation is not perfect, and more optimized filtering strategies than
spin-echo must be considered.
Concatenated sequences like Carr-Purcell-Meiboom-Gill (CPMG) are a generalization
of spin-echo~\cite{cywinski_how_2008}
with $2nX$ pulses separated by an interval $t/(2n)$.
This filters out frequency components
smaller than $2n/t$, therefore the performance of DD increases with $n$.

\textbf{Dynamical decoupling with finite pulse durations---}
In the context of our model, the $X$ pulses always have
 a finite (non-zero) duration $t_\pi$ that we need to take into account to include
 DD sequences at the pulse-level.
If $2t_\pi\le t$, we can realize a spin echo sequence as described above.

To realize an optimized concatenated CPMG sequence, i.e., with the largest
possible value of $n$, let us make the following observation.
If we try to fit a maximized number of
$2nX$ pulses with given duration $t_\pi$ in a finite $t$ window,
this results in a sequence where
the effective idle duration becomes negligible.
In other words, an optimized CPMG sequence with a finite pulse duration can be seen as a
single $X^{2n}=R_X(2n\pi)$ rotation, which will be relatively easy to describe
at the pulse-level.

\subsubsection{Gate exchange noise}
So far, we only considered noise affecting the qubit frequency.
To model noisy quantum circuits more accurately, one needs to take 
into account fluctuations of the coupling constants of the Heisenberg
Hamiltonian~\cite{gravier_simulated_2025}.
To do so, we replace
$J_{i,j}(t)$ in Eq.~\eqref{eq:heisenberg} by $J_{i,j}(t)+\delta J_{i,j}(t)$,
where $J_{i,j}(t)$ still refers to the ideal pulse, and
\begin{align}
  \delta J_{i,j}(t)
  &= J_{i,j}(t)\left(e^{\eta_{i,j}(t)}-1\right)
  \nonumber \\
  &\approx
  \frac{J_{i,j}(t)}{J}
  \tilde{\epsilon}_{i,j}(t), \label{eq:J_distort}
\end{align}

\noindent where $\tilde{\epsilon}_{i,j}(t)=J\eta_{i,j}(t)$ represents a small noisy field that originates from the
fluctuations of gate-exchange potentials, and $J$ represents the 
maximum value of $J(t)$.

Now we provide a physical example that will help us
parameterizing
the noisy time traces $\tilde{\epsilon}_{i,j}(t)$
in terms of a coherence time $T_J^*$, in a way that
is analog to our description of the qubit frequency shifts $\epsilon_{i}(t)$.

Under the adiabatic evolution described in Sec.~\ref{sec:twoq_gate},
the presence of a noise term leads to a gate $R_{ZZ}(\theta+\Delta\theta)$, with
\begin{equation}
\Delta \theta = \int_0^t \frac{J_{i,j}(t')}{J}\tilde{\epsilon}_{i,j}(t')dt'
\approx \int_0^t\tilde{\epsilon}_{i,j}(t')dt',
\end{equation}
where in the second equality we assume the coupling $J_{i,j}(t')$ is set
to its maximal value $J$
during most of the evolution time, i.e., we consider a square pulse and neglect
the effect of the adiabatic ramping of the coupling\footnote{A better approximation could be to replace $J_{i,j}(t)$ by its
time-averaged value.}.

We obtain a quantum channel
\begin{align}
  \mathcal{E}_{ZZ}(\rho)
  &=
    \mathbb{E}
    \left[
      R_{ZZ}(\theta+\Delta\theta)
      \rho
       R_{ZZ}^\dag(\theta+\Delta\theta)
    \right]
    \nonumber \\
    &=
  R_{ZZ}(\theta)
  \mathcal{E}(\rho)
R_{ZZ}^\dag(\theta),
 \end{align}
with
 \begin{align}
    \mathcal{E} (\rho)
    &= \frac{1+\tilde{C}(t)}{2}\rho
    +\frac{1-\tilde{C}(t)}{2}ZZ\rho ZZ,\end{align}
and the contrast $\tilde C(t)=\mathbb{E}
\left[\exp
  \left(i\Delta\theta\right)
 \right]$.

 Summarizing, just as in the case of qubit noise frequency,
 the effect of gate-exchange noise can be related to a dimensionless contrast quantity
 $\tilde C(t)$.
 Reproducing the analytical calculations given in Sec.~\ref{sec:noise_types},
 this allows us to define a coherence time $T_J^*$ through
the decay of
 $\tilde C(t)$.
As a consequence, and as described below, we can parametrize the time traces
 $\tilde \epsilon_{i,j}$ unambiguously in our code API,  by using a coherence time $T_J^*$ as input parameter for the three noise types.

\section{Description of the API}\label{sec:description}

Our package realizes the steps (1) to (3) of simulation
described in Sec.~\ref{sec:organization} in a modular way. 
This approach allows users to carry out each step themselves 
and visualize the corresponding circuits.

\subsection{Installation and documentation}

Our package can be installed via pip
\begin{lstlisting}
  pip install spin-pulse
\end{lstlisting}
The source code is available on
 \href{https://github.com/quobly-sw/Spin-Pulse}{Github}, and 
 the API documentation is available 
  \href{https://quobly-sw.github.io/SpinPulse}{here}.

\subsection{Hardware specifications}
To begin with, the user is prompted 
to specify the free parameters of the model in Sec.~\ref{sec:model}.
These parameters are incorporated in an instance of the class
\texttt{HardwareSpecs}, which will be 
used through the transpilation
process. For example, the code
\begin{lstlisting}[language=Python]
  num_qubits = 3
  B_field,delta,J_coupling = 0.3,0.3,0.03
  ramp_duration = 5
  specs = HardwareSpecs(num_qubits,B_field,delta,J_coupling,Shape.GAUSSIAN,ramp_duration)
\end{lstlisting}
instantiates a quantum computer model with \texttt{num\_qubits=3} qubits and
coupling constants \texttt{B\_field}, \texttt{delta}, \texttt{J\_coupling} that
correspond to the maximum value authorized by the hardware for the fields $B_i(t)$,
$\delta \omega_i(t)$, and $J_{i,j}(t)$ in the model.
The last two arguments refer to the shape of the
 pulses, here Gaussian, used to realize
the gates, as well as the time needed to reach the maximal value
for the different fields or coupling. 
Finally, in the present implementation of the package, we assume a one-dimensional
connectivity of the quantum computer, i.e., there is a native two-qubit gate interaction
for each pair $(i,i+1)$ of qubits.

We remind that the times are expressed in units of the time resolution $t_s$, and
the angular frequencies as rad$\cdot t_s^{-1}$.

\subsection{Gate transpilation}
Once the hardware specifications are set, we are ready to tackle the gate transpilation
step of
our simulation tool. This begins by specifying a circuit in \texttt{qiskit} language
\begin{lstlisting}[language=Python]
qreg = qi.QuantumRegister(num_qubits)
circ = qi.QuantumCircuit(qreg)
circ.cx(0,1)
\end{lstlisting}
Here we create a simplistic circuit with a \texttt{CNOT} gate between the first two
qubits, while the last qubit remains idle.
If one is interested in using our library for a circuit written in another language,
such as \texttt{Pennylane}, or $\texttt{Cirq}$, one can use the module \texttt{qiskit.qasm3}
to perform the conversion via the OpenQASM language.

The gate transpilation consists in converting these circuits into the native gate
set of our model. To achieve this, the class \texttt{HardwareSpecs} is equipped
with two parametrized \texttt{qiskit} transpilation pass managers.
First, we realize a standard transpilation into our gate set, and then we split the
$R_{ZZ}$ gates into two parts, following Eq.~\eqref{eq:rzz}. The method
\texttt{gate\_transpile} performs these two passes
\begin{lstlisting}[language=Python]
isa_circ = specs.gate_transpile(circ)
\end{lstlisting}
  Here, the object \texttt{isa\_circ} is an Instruction Set Architecture (ISA)
  circuit, meaning that it can
  be physically implemented in our model.
   At this stage however, we did not yet specify
 how such gates will be generated at the pulse-level, which is the purpose of the pulse
  transpilation step.
  Also note that \texttt{qiskit} uses initial and final layouts to map
  the initial and final qubit states of the input circuit to the qubits of the ISA circuit.
  This crucial information will be preserved via dedicated class attributes
  during the different transpilation and simulation steps.

\subsection{Pulse transpilation}
Our gate set is composed of four parametrized gates $R_X$, $R_Y$, 
$R_Z$, $R_{ZZ}$, whose angle
$\theta$ is simply related to the time integral of their respective generating fields $B_i(t),\,\delta \omega_i(t),\, J_{i,j}(t)$.
The problem of implementing a gate with  a certain rotation angle is reduced to
describing pulses with well-defined integrals, under the constraints of the maximal
coupling strengths specified above.

One further requirement to describe accurately the pulses is the assignation of starting and ending times for all quantum gates in a consistent manner, while
referencing the time windows during which certain qubits are idle (not subject to any pulse).

Pulse transpilation is organized via the class \texttt{PulseCircuit}, whose constructors
return a finalized pulse-level description of the circuit. This construction
operates as follows:

\textbf{(1) Circuit into layers---}
 An ISA circuit is decomposed into a list of layers $L_1,\dots,L_m$, which correspond to portions
of the circuit where each qubit is involved at most in one quantum gate. Each layer
corresponds to an instance \texttt{pulse\_layer} of the \texttt{PulseLayer} class, which contains all the pulses needed to
implement this specific layer.

\textbf{(2) Layers into sequences---} Each \texttt{pulse\_layer} is assembled
as two lists of \texttt{PulseSequence} objects acting
on one and two-qubits, respectively.
This class \texttt{PulseSequence} describes the set of pulses corresponding to single- and two-body gates
during a layer.
By definition, all \texttt{PulseSequence} instances of a \texttt{PulseLayer}
have the same duration, which is an integer attribute in both classes.

\textbf{(3) Sequences into instructions---} Finally the \texttt{PulseSequence} objects are made of time ordered sequences of
\texttt{PulseIntruction} objects, which represent a bare pulse-level instruction, in the sense that
the corresponding Hamiltonian evolution $h_i(t)$ or $h_{i,j}(t)$ is governed by a single time-dependent function.
 In order to be able to implement both gates
 and idle evolutions,
\texttt{PulseIntruction} has two subclasses: \texttt{RotationInstruction}, and
\texttt{IdleInstruction}.
The class \texttt{RotationInstruction} is typically initialized with the
method \texttt{from\_angle}, which adjusts the amplitude
and duration of the pulse once its shape is fixed; this guarantees that a certain rotation angle $\theta$ is achieved.

The three steps of pulse transpilation are realized via the constructor \texttt{from\_circuit}
 of the class \texttt{PulseCircuit}:
\begin{lstlisting}[language=Python]
pulse_circ = PulseCircuit.from_circuit(isa_circ,specs)
pulse_circ.plot()
\end{lstlisting}
where the \texttt{plot} method allows for visualizing the complete pulse scheduling of the circuit.

The use and importance of the three pulse scheduling classes, \texttt{PulseLayer}, \texttt{PulseSequence} and \texttt{PulseInstruction}, is illustrated in Fig.~\ref{fig:package}
for a small quantum circuit example.
The corresponding graphical representation of the \texttt{PulseCircuit}
 object was obtained with the method \texttt{plot}.
 The first layer contains three \texttt{PulseSequence} instances,
 which are composed of a \texttt{RotationInstruction} for qubit 1, a 
\texttt{RotationInstruction} followed by an  \texttt{IdleInstruction} for qubit 2, 
and an \texttt{IdleInstruction} for qubit 3.
 In the second layer, a \texttt{PulseSequence} acts on two qubits (represented on the intermediate row between them), in order to realize an
$R_{ZZ}$ gate; it is made of two \texttt{IdleInstruction} instances and a \texttt{RotationInstruction}.
During the two idle periods, the frequency of the two qubits
switches between $0$ and a large plateau value, which is described by single-qubit 
\texttt{PulseSequence} instances.
The overall sequence realizes the (quasi-) adiabatic Ising interaction described
in the model section, where the parameter $\Delta(t)$ is set to a plateau value, before the Heisenberg interactions
begin.

\subsection{Numerical integration}
Numerical integration consists in 
taking a \texttt{PulseCircuit} object and converting it back into
a gate circuit, i.e., calculating the evolution operator
\begin{align}
u = e^{-ih(t_f)}\dots e^{-ih(t_i)} 
\end{align}
associated with each time-dependent Hamiltonian acting on one or two qubits.
This is done with
\begin{lstlisting}[language=Python]
applied_circ = pulse_circ.to_circuit()
\end{lstlisting}
where \texttt{applied\_circ} is a \texttt{qiskit} circuit which
can be represented graphically, as in Fig.~\ref{fig:package},
but also simulated numerically via \texttt{qiskit\_aer} or another simulator.

In the absence of noise, the ISA circuit \texttt{isa\_circ} and the final circuit 
\texttt{applied\_circ}
only differ due to the fact that the SW adiabatic
transformation is an approximation.
Indeed, the implemented circuit integrates the Heisenberg
 interactions Eq.~\eqref{eq:heisenberg}, and therefore captures the non-perturbative, non-adiabatic effects.
This can be quantified by calculating the process fidelity between the evolution operators of the two circuits
\begin{lstlisting}[language=Python]
print(pulse_circuit.fidelity())
0.9999992514359283
\end{lstlisting}
Note that the process fidelity can be further increased by reducing the 
coupling strength $J$, at the expense of also increasing the circuit duration.

\begin{figure*}
  \includegraphics[width=\textwidth]{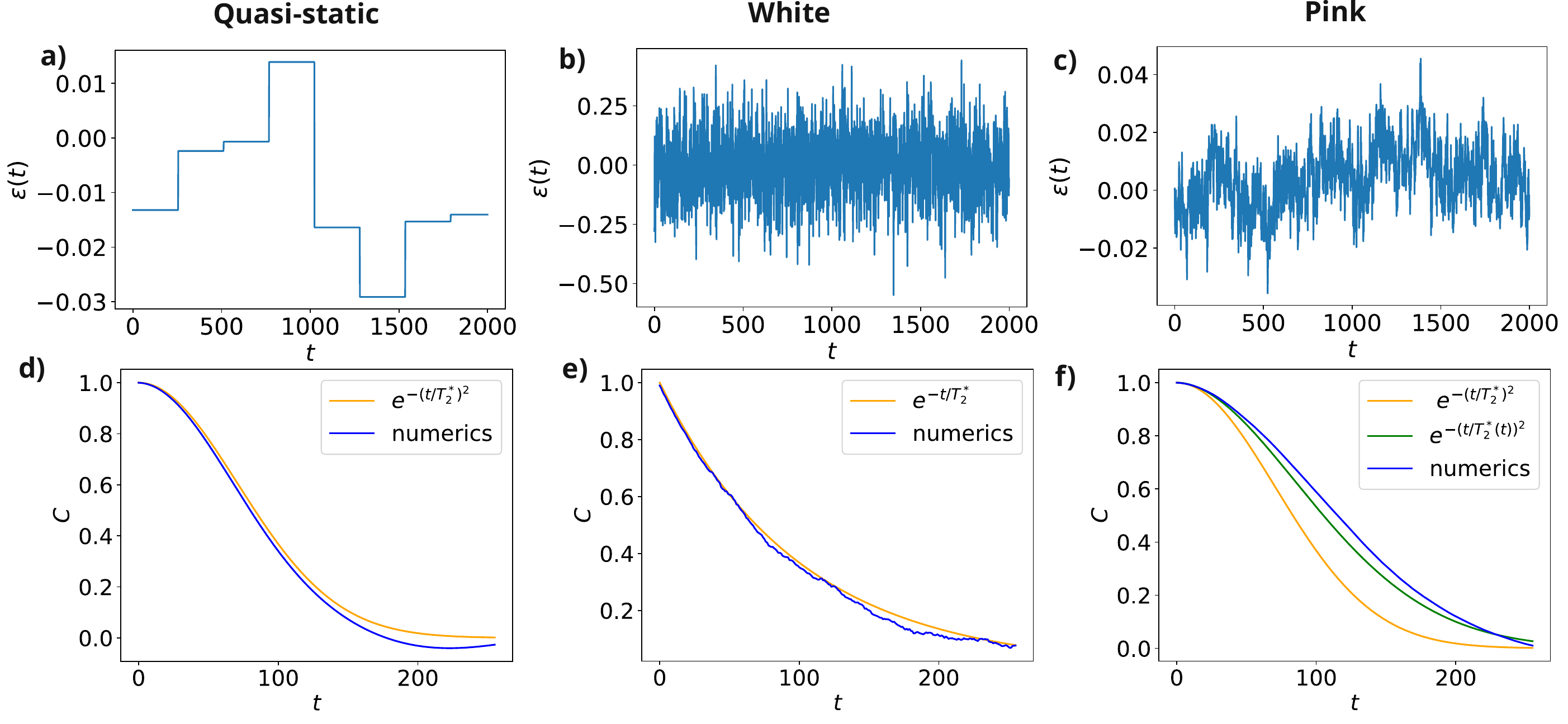}
  \caption{Examples of time traces $\epsilon(t)$ for a) quasi-static,
  b) white, and c) pink noise, for the
  same value of $T_2^*=100$, duration $2^{18}$. For quasi-static noise, we use
  time traces of $\texttt{segment\_duration}=250$.
  In panels d),e), f), we show the corresponding
  averaged Ramsey contrast calculated using the
   method
  \texttt{plot\_ramsey\_contrast}.
  }
  In panel f), we show two different analytical curves: In orange, 
  the expression of the Ramsey contrast $C(t)=\exp\big(-t^2/(T_2^*(t)^2~\big)$ with the time-dependent value $T_2^*(t)$
  given in Eq.~\eqref{eq:T2vst}, and in blue the expression obtained with the approximation $T_2^*(t)\approx T_2^*(1)$. 
  \label{fig:timetraces}
  \end{figure*}

\subsection{Integration of the noise model}

Our Sec.~\ref{sec:model} describes a model with
noise affecting the qubits in the form
of random time traces $\epsilon_i(t)$,  $\tilde{\epsilon}_i(t)$. To quantify ensemble-averaged quantities,
 such as fidelities and
quantum channels, we introduce now the \texttt{ExperimentalEnvironment}
 class.
This object references random realizations $\epsilon_i(t)$, $\tilde{\epsilon}_i(t)$
of our noise model.
Over the physical time span of an \texttt{ExperimentalEnvironment} instance,
we simulate sequentially multiple random realizations of a \texttt{PulseCircuit},
in order to extract average quantities as in a true experimental scenario.
The \texttt{duration} attribute of an \texttt{ExperimentalEnvironment} is therefore taken
typically much longer than that of the \texttt{PulseCircuit} we are interested in.

An \texttt{ExperimentalEnvironment} instance can be initialized as
\begin{lstlisting}[language=Python]
exp_env = ExperimentalEnvironment(hardware_specs=specs,noise_type=NoiseType.PINK,T2S=500,duration=2**13,segment_duration=2**13)
\end{lstlisting}
admitting the three types of noise \texttt{QUASISTATIC}, \texttt{WHITE},
 and \texttt{PINK}.
The parameter \texttt{T2S} defines the strength of the time traces
$\epsilon_i(t)$ according to the expressions derived
in Sec.~\ref{sec:model}.

In addition to the \texttt{duration} parameter, 
the integer \texttt{segment\_duration} parameter plays a specific role
 depending on the noise type. 
For quasi-static noise, the signal is constant 
over segments of duration \texttt{segment\_duration}.
In the opposite limit of white noise, we enforce \texttt{segment\_duration = 1}. Finally, for pink noise, it provides a knob
 to set a low frequency cutoff 
 $f_{\min} = (\text{segment\_duration})^{-1}$
  to the generated time traces. 

An example of time trace $\epsilon(t)$ for a single qubit is shown in Fig.~\ref{fig:timetraces}a)-c) for the
three types of noise. In panels d)-f), we show the corresponding Ramsey
contrasts, calculated numerically and averaged over the time trace durations,
together with the results from the analytical expressions.

The \texttt{ExperimentalEnvironment} class also admits an optional \texttt{TJS} parameter that  specifies the exchange noise. As for the case of the qubit frequency noise $\epsilon_i(t)$, the code samples
time traces $\tilde{\epsilon}_{i,j}(t)$ with coherence time \texttt{TJS}, following the analytical
expression of the contrast given in Sec.~\ref{sec:model}. Together with the value of $J$,
we obtain the distortion factor $\eta_{i,j}(t)$ that alters the coupling
in the linear approximation \eqref{eq:J_distort}.

\subsubsection{Noise-accurate simulations}

We are now ready to discuss how to realize noise-accurate
simulations. This is done by attaching an experimental environment instance to a pulse circuit.
\begin{lstlisting}[language=Python]
pulse_circuit = PulseCircuit.from_circuit(isa_circ,specs,exp_env=exp_env)
\end{lstlisting}
Internally, our package adds  a \texttt{time\_trace} attribute to each one qubit \texttt{PulseSequence}; it represents the values of $\epsilon_i(t)$ in the time window
$t\in [t_\mathrm{lab}-\Delta t,t_\mathrm{lab}]$
where the circuit is executed. Here $t_\mathrm{lab}$ represents the ending time
of the circuit in the laboratory frame, and $\Delta t$ the circuit time duration.
During numerical integration, the time trace contributes to the total Hamiltonian describing the
system and results in a noisy version of the gates.

{\it Computing average circuit metrics---}
In order to perform an ensemble-average $\mathbb{E}$ over the circuit realizations $u$, and extract for instance the gate fidelity
$\mathcal{F}$ defined in Eq.~\eqref{eq:gate_fidelity}, the class \texttt{PulseCircuit} provides a method \texttt{assign\_time\_trace} that shifts
the laboratory time $t_\mathrm{lab}$ associated with an experimental environment
by an increment $\Delta t$,
and leads to shifted time traces $\epsilon_i(t)$.
Repeating the procedure multiple times
gives access to ensemble-averaged quantities.
In particular, one can access the fidelity $\mathcal{F}$ using
\begin{lstlisting}[language=Python]
  pulse_circuit_noise.mean_fidelity(exp_env)
  0.9727939521865222
\end{lstlisting}
Similarly, the method \texttt{mean\_channel} gives the average noisy quantum
 channel in
the superoperator representation $\mathcal{S}$.
Any other averaged quantity can be accessed using the method
\texttt{averaging\_over\_samples}.

{\it Simulating an experimental procedure with measurements---}
In an experiment, one realizes a random instance of the circuit
followed by a measurement of a bitstring $s=s_1,\dots,s_N$.
According to Born's rules, the bitstring $s$ is sampled 
from the distribution $|\braket{s|u\ket{0^{\otimes N}}}|^2$.

We provide a function \texttt{get\_bitstrings} to repeat this 
procedure iteratively over multiple realizations of noisy circuits $u$, and for the duration of an \texttt{ExperimentalEnvironment}.
\begin{lstlisting}[language=Python]
pulse_circuit_noise.run_experiment(exp_env)
\end{lstlisting}
returning a dictionnary of bitstrings counts.

\subsection{Dynamical decoupling}

\begin{figure}
  \begin{center}
  \includegraphics[width=\columnwidth]{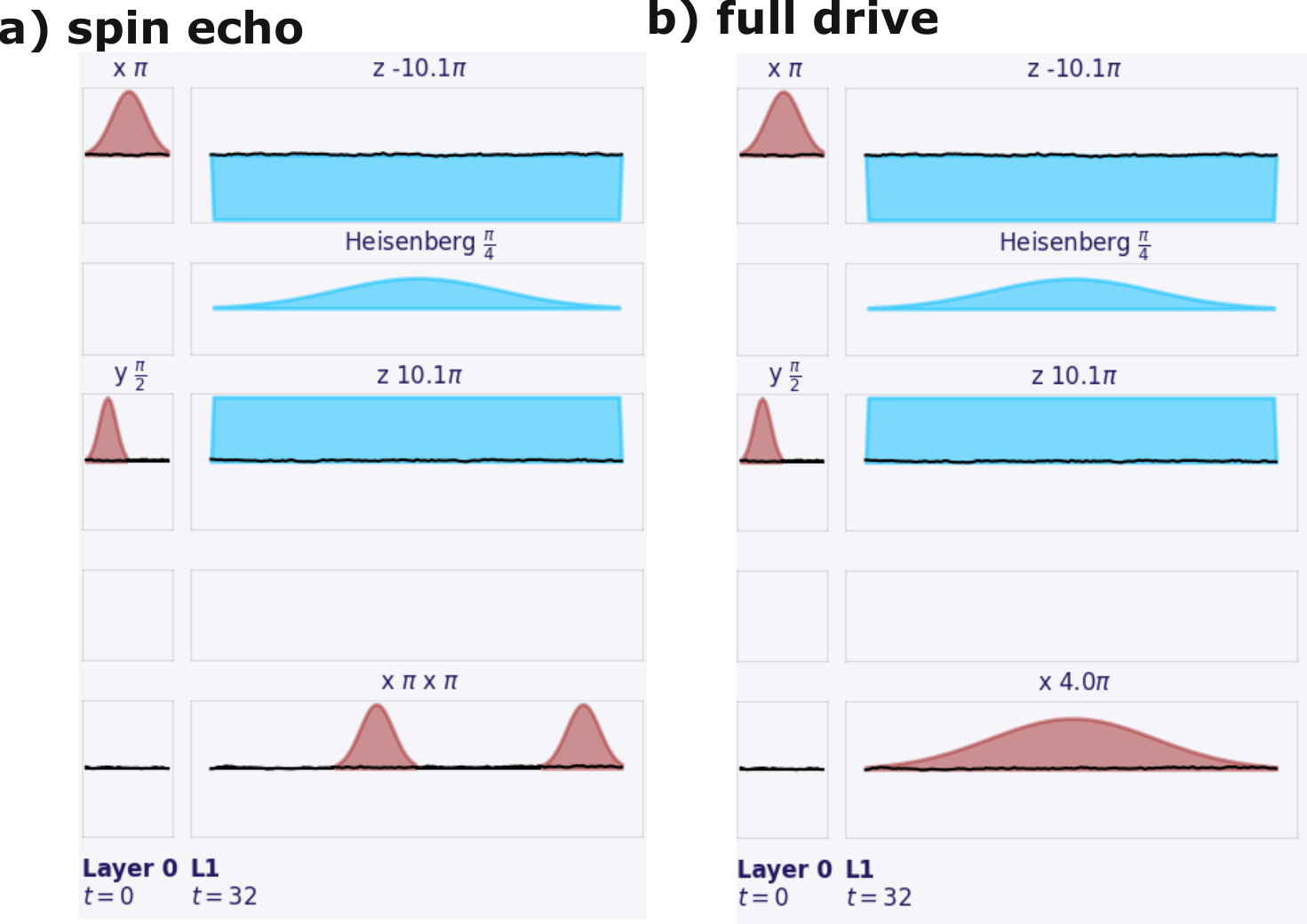}
  \end{center}
  \caption{
    Dynamical decoupling for the first three layers of the circuit
     shown in  Fig.~\ref{fig:package}.
    With a) spin echo, only the second idle pulse sequence can accommodate the
    two $X$ pulses. With b) full drive, this is replaced by an $R_{X}(2n\pi)$ rotation
    whose amplitude is adjusted to occupy the full duration of the sequence. Here, $n=1$.
  }
  \label{fig:dynamical_decoupling}
\end{figure}

The \texttt{HardwareSpecs} class takes an optional argument
\texttt{dynamical\_decoupling} which can be
\texttt{SPIN\_ECHO} or \texttt{FULL\_DRIVE}, based on the
sequences given in Sec.~\ref{sec:model}.
During the pulse transpilation, idle time is measured for each \texttt{PulseSequence}
instance, and converted to a dynamical decoupling sequence.
For spin echo, the package checks that two $X$ pulses can be fitted in the idle time.
Similarly, for full-drive, the largest possible $n$ is used to achieve
an $R_X(2n\pi)$ rotation.
This procedure is illustrated in Fig.~\ref{fig:dynamical_decoupling}.

 \begin{figure*}
    \includegraphics[width=\textwidth]{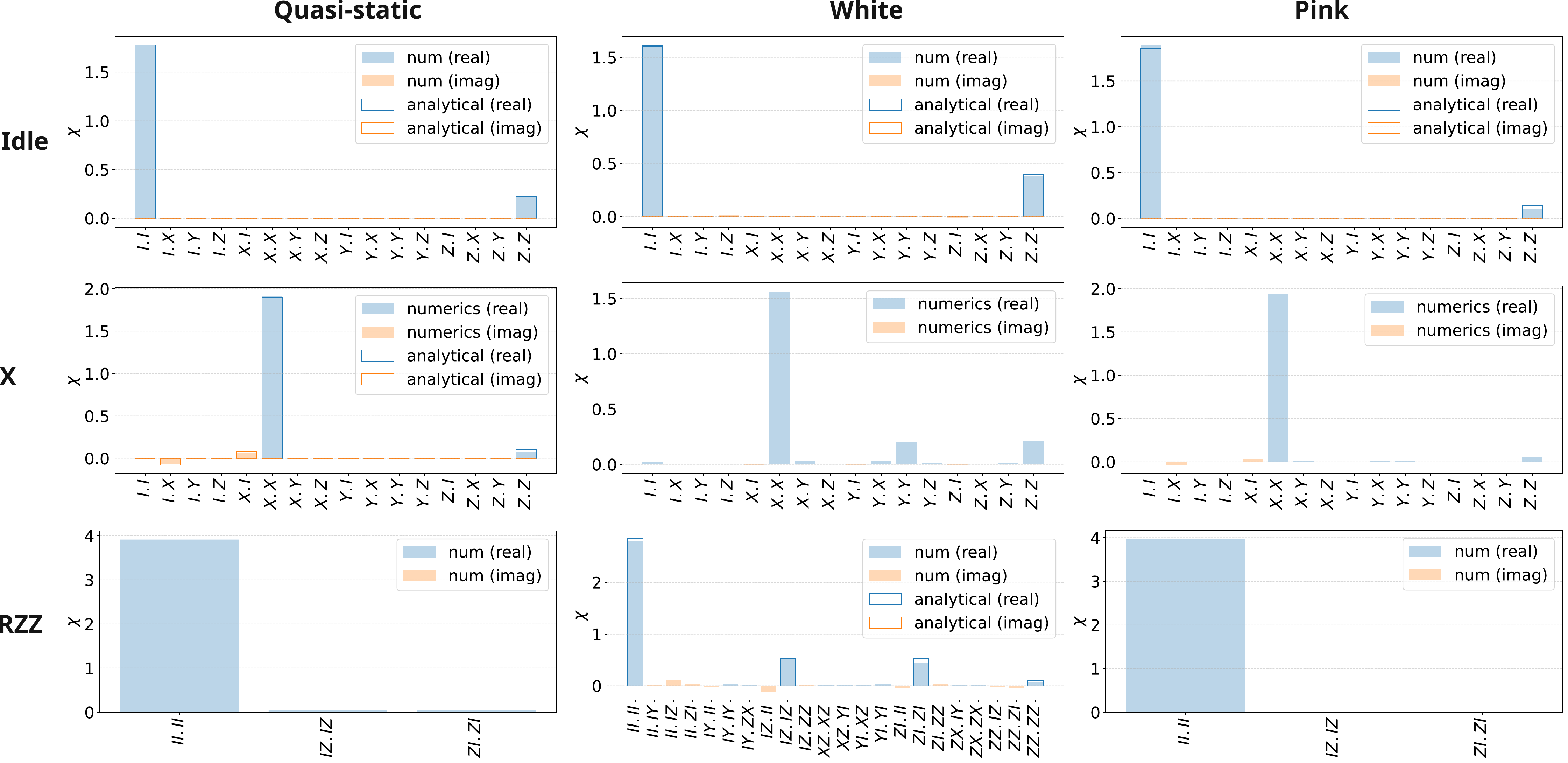}
    \caption{Gate channels in the $\chi$ matrix form for idle
     (top), $X$ (middle) and $R_{ZZ}$ (bottom) gates
    as shown using the method \texttt{plot\_chi\_matrix}. 
    We use the three types of noise,
    with the (many) parameters shown in the file
     \texttt{AverageSuperoperatorsNoisyGates.py}. The analytical
    expressions are written in the model section \ref{sec:model}.}
    \label{fig:gatechannels}
    \end{figure*}

  \section{Examples of advanced usage noise-accurate circuit simulations}\label{sec:advanced}

We provide in our package seven notebooks illustrating, step-by-step,
practical use cases. In this article, we explain in details two of them, 
and refers the reader to our documentation for information about the other notebooks.

  \subsection{Quantum gate channels under noise}

We first illustrate our package for evaluating
the effect of noise on our quantum gates.
The corresponding parameters and codes are presented in the \texttt{jupytext} notebook
\texttt{AverageSuperoperatorsNoisyGates.py}.

We consider three types of gates (idle, $X$ and $R_{ZZ}$), and three types of noise (quasi-static, white and pink). We numerically evaluate 
the gate channels
$\mathcal{S}$ using the method \texttt{mean\_channel} of the class
\texttt{PulseCircuit} and then plot the corresponding $\chi$ matrix using the function
\texttt{plot\_chi\_matrix}. It also plots the analytical expressions of the
channels derived in the model introduced in Sec.~\ref{sec:model} when available.
The results are shown in Fig.~\ref{fig:gatechannels}.

For an idle gate, we obtain as expected that the channel has two dominant contributions,
$I\cdot I$ and $Z\cdot Z$, the latter representing the probability to be affected by
a phase-flip event. The magnitude of this probability agrees with our analytical expressions,
providing a sanity check for the numerical implementation of the noise model.

For the $X$ gate under quasi-static noise,
we observe an agreement with our analytical result showing a $Z\cdot Z$ contribution.
 We observe a similar behavior for
pink noise, while for white noise there is an additional $Y\cdot Y$ contribution.
Having in mind for example quantum error mitigation and correction
strategies for biased noise, this shows that our package is useful to assess
realistically how the various noise types may lead to different quantum channels.

Finally, for the case of the $R_{ZZ}$ gate, we confirm our analytical observation that
the gate is mainly affected by a spin-echo filtered version of the noise, being essentially
perfect for quasi-static and pink noise, and in perfect agreement with our analytical
estimation for white noise.

    \subsection{Tensor-network pulse-level simulations using \texttt{quimb}}

  The steps (1) to (3) of our package end up with an instance 
    of a \texttt{qiskit} circuit, which is a data structure representing
  all the gates of the circuits as unitary matrices (as illustrated
  in the top right panel in Fig.~\ref{fig:package}).
  We illustrated the calculation of the fidelity above, using \texttt{qiskit} functions.
 This is however limited to moderate system sizes, as it
 requires
  the evaluation of the total evolution operator of the circuit, an exponentially large object in the number of qubits.

Tensor-network methods allow for extending the scope of classical simulations
of quantum circuits, and are available from a vast variety of
open-source packages. As these methods require specifically a
quantum circuit as input, they can be used directly in the context of our
package.

For instance, we provide the translating function
\texttt{qiskit\_to\_quimb}
that recasts a \texttt{qiskit} circuit
into the \texttt{CircuitMPS} format of the \text{quimb} tensor-network
 library~\cite{gray_quimb_2018}.
 This only
consists in reformatting the unitary matrices of the \texttt{qiskit} circuit,
and can be done easily for any gate-based simulator.

\begin{figure}
\includegraphics[width=\columnwidth]{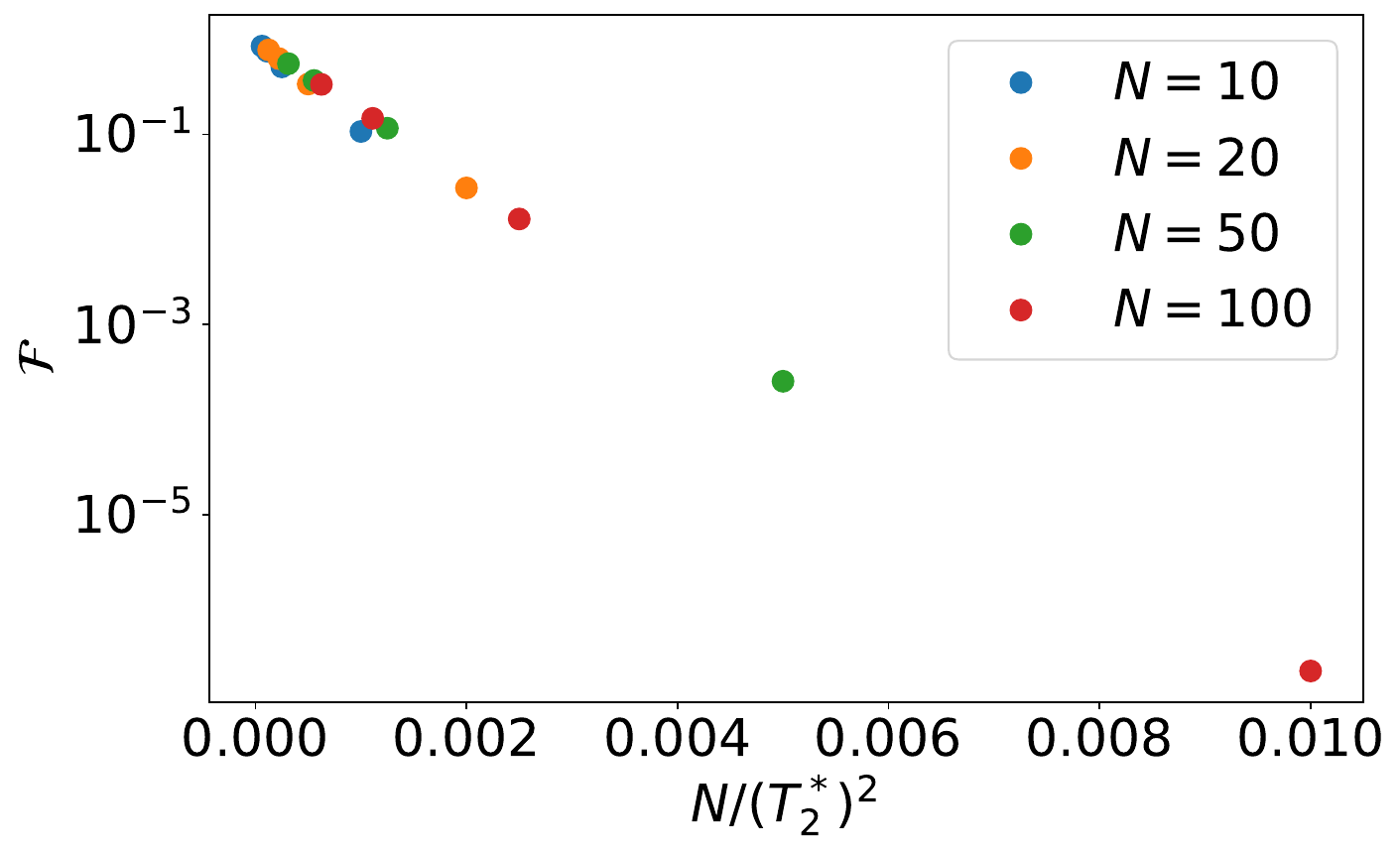}
\caption{State fidelity for a cluster state prepared within our model,
in the presence of pink noise of coherence time $T_2^*$ and for various qubit numbers $N$. All the other simulation
parameters are presented in the dedicated Jupyter notebook.}
\label{fig:quimb}
\end{figure}

We show in Fig.~\ref{fig:quimb} the state-fidelity for the example of a uni-dimensional
 cluster state for up to $N=100$ qubits.
Such large system size is accessible because the preparation circuit has a depth of two (one layer of Hadamard gate,
one layer of $CZ$ gates). Therefore, the calculation of the matrix-product-state representing
the wave-function and of the fidelity with respect to the ideal state
requires a computation time that is linear in $N$~\cite{gray_quimb_2018}.
Note that the observed dependence $\mathcal{F}\sim \exp\big(-\alpha N/(T_2^*)^2\big)$
resembles the form of fidelity measurements $\mathcal{\tilde{F}}=\exp(-\epsilon_B N D/2)$
observed in superconducting qubit experiments~\cite{arute_quantum_2019}, with $\mathcal{\tilde{F}}$ the
cross-entropy fidelity, $D$ the depth, and
$\epsilon_B$ a single-qubit error rate.

\section{Conclusion}
 We present a modular package that implements a quantum computing model
 relevant for spin qubit architectures.
 This can be used as a pulse-level simulator to predict for instance quantum gates fidelities or to perform noise-accurate simulations
 of large-scale quantum experiments.

Our model will to evolve in various ways,
in order to follow
 both hardware and software (e.g. gate implementation)
  improvements.
 Our package will incorporate these changes through regular updates.
 First, pulse shaping and optimal control strategies will be incorporated naturally
 in our package.
 This will happen by defining suitable \texttt{PulseInstructions} subclasses that can be optimized
 in closed loops using our fidelity estimation methods, in the spirit of what is
 available in \texttt{qutip}~\cite{lambert_qutip_2024} for example.
 We will also enrich our connectivity from one to two-dimensional layouts, and 
 extend our native gate set using e.g., exchange-type operations
~\cite{burkard_semiconductor_2023}.

 Another goal is to make our package one step closer
  to a microscopic description of our devices.
 For instance, in a future release, we may  include the scheduling constraints
  in driving
  the qubits by electron spin resonance  (ESR) drives~\cite{burkard_semiconductor_2023},
  resulting in a \texttt{PulseCircuit} class that will mimic more closely what happens
  in the actual hardware. This will, in turn, drive new theory insights on how
  to operate spin qubits in large-scale quantum computers.

\section*{Acknowledgments}

$\quad$We thank Jean-Marc Volle and Alain Champenois for useful discussions.

Work in Quobly is funded by the French National Research 
Agency via the research programs Plan France 2030
 HQI (ANR-22-PNCQ-0002), 
and by the European Union’s Horizon 2020 research and innovation program, under grant agreement No951852 (QLSI project).
VS acknowledges funding from the Plan France 2030 through grant PRESQUILE (ANR-22PETQ-0002)
\bibliographystyle{quantum}
\bibliography{refs.bib}

\appendix
\onecolumn

\section{Adiabatic Schrieffer-Wolff transformation}
\label{app:adiabatic}
In this appendix, we explain how to derive the evolution
operator for a time-dependent Hamiltonian expressed
via Schrieffer-Wolff transformation
\begin{equation}
h_\mathrm{SW}(t) = e^{S(t)}h_\mathrm{d}(t)e^{-S(t)},
\end{equation}
where we assumed that $h_\mathrm{d}(t)$ is a diagonal Hamiltonian in the configuration basis $\{s\}$
\begin{equation}
h_\mathrm{d}(t) = \sum_{s}c_s(t)\ket{s}\bra{s}.
\end{equation}
As $e^{S(t)}$ is a unitary matrix, we also note that the numbers $c_s(t)$ represent
the instantaneous eigenenergies of $h_\mathrm{SW}(t)$.

Let us further assume that
at $t=0$, we have $S(0)=0$.
In our two-qubit model, this is satisfied by setting the
coupling $J(0)=0$. This implies that $h_\mathrm{SW}(0)=h_\mathrm{d}(0)$ and
thus that the initial Hamiltonian is diagonal
in the configuration basis.
Consider an initial wavefunction in this basis
\begin{equation}
\ket{\psi(0)} = \sum_s \psi_s \ket{s}.
\end{equation}
According to the quantum adiabatic theorem, each amplitude will pick up an adiabatic phase associated with the integral of each instantaneous energy eigenstate $c_s(t)$, i.e.,
\begin{equation}
\ket{\psi(t)}=\sum_s  \exp\left(-i \int_0^t c_s(t')\right)\psi_s \ket{s}
= \mathcal{T}\exp\left(-i \int_0^t h_\mathrm{d}(t')dt'\right) \ket{\psi(0)}\,.
\end{equation}
Therefore, the evolution operator is given by
\begin{equation}
u(t) = \mathcal{T}\exp\left(-i \int_0^t h_\mathrm{d}(t')dt'\right)\, .
\end{equation}

\section{Details on noisy  gates calculations}\label{app:noise_channels}

 In this appendix, we give additional details on the analytical calculation of noisy $X$ and $R_{ZZ}$ presented in the main text.

\subsection{Noisy $X$ gates}

We start from the generating Hamiltonian
\begin{equation}
  h_i = \frac{B_0}{2}X+\frac{\epsilon}{2}Z=\frac{\Omega}{2} P,
  \end{equation}
  with $\Omega=\sqrt{B_0^2+\epsilon^2}$, and $P=\cos(\theta) X+\sin(\theta) Z$,
  $\cos(\theta)=B_0/\Omega$, $\sin(\theta)=\epsilon/\Omega$. 
  We consider the quasi-static noise model with $\mathbb{E}[\epsilon]=0$, 
  and $\mathbb{E}[\epsilon^2]=\sigma^2$.
  It gives the evolution operator
  \begin{equation}
  u
  =\cos\left(\frac{\Omega}{2} t\right)I-i\sin\left(\frac{\Omega}{2} t\right) P\,.
  \end{equation}

  Let us now consider that we want to implement $X=R_X(\pi)$.
  This implies $B_0=\pi/t$.
We calculate the effect of noise up to second order in $\epsilon/B_0$:
  \begin{align}
    \Omega&= B_0\Bigg(1+\frac{\epsilon^2}{2B_0^2}\Bigg)\nonumber \\
    \cos\theta &= 1-\frac{\epsilon^2}{2B_0^2}\nonumber \\
    \sin\theta &= \frac{\epsilon}{B_0}\nonumber \\
  \frac{\Omega t}{2}&= \frac{\pi}{2}\Bigg(1+\frac{\epsilon^2}{2B_0^2}\Bigg) \nonumber \\
  \cos\left(\frac{\Omega}{2} t\right) &= -\frac{\pi\epsilon^2}{4B_0^2} \nonumber \\
  \sin\left(\frac{\Omega}{2} t\right) &= 1\, ,
  \end{align}

  \begin{equation}
  u_\pi= -\frac{\pi\epsilon^2}{4B_0^2}I
  -i \left(
  1-\frac{\epsilon^2}{2B_0^2}
  \right)X
  -i \frac{\epsilon}{B_0}Z,
  \end{equation}
which gives an average channel

\begin{equation}
  \mathcal{E}_{\pi}(\rho)
  =
  \left(
  1-\frac{\sigma^2}{B_0^2}
  \right)
  X \rho X
  +i\frac{\pi \sigma^2}{4B_0^2} (I\rho X-X\rho I)
  +
  \frac{\sigma^2}{B_0^2}Z\rho Z\,.
  \end{equation}

\subsection{Noisy $R_{ZZ}$ gates}

\subsubsection{General formulation}
Let us continue the calculation shown in Ref.~\ref{sec:model}, by adding
a noise Hamiltonian $\delta h_i(t)+\delta h_j(t)$
to the noiseless part $h_{i,j}(t)$.

The Schrieffer-Wolff (SW) transformation remains valid in this case, with
the replacements
\begin{equation}
  \delta \omega_{i,j}(t)
  \to
  \delta \omega_{i,j}(t)
  +\epsilon_{i,j}(t)
\end{equation}
and we obtain, as before,
\begin{align}
  u(t)&=\mathcal{T}\exp\left(-i\int_0^t h_0(t')dt'\right)
  u_\mathrm{sh}(t)
  \nonumber \\
  &=R_{ZZ}\big(\theta(t)\big)
  v(t).
 \end{align}
 The difference with the noiseless case arises when considering the spin-echo
 trick
 \begin{align}
  \tilde u(t)
  =(u(t/2)X_iX_ju'(t/2)X_iX_j)=R_{ZZ}\big(\theta(t)\big)v(t/2)[v'(t/2)]^\dag,
 \end{align}
 where $v(t/2)$ and $v'(t/2)$ differ due the fact the noise terms
 $\epsilon_i(t)$, $\epsilon_j(t)$ may vary during the two consecutive
 implementations of the gate. We assumed that the $X$ gates are not affected
 by the noise.
 Considering in addition in first approximation that the Stark-Shift terms
  $h_\mathrm{sh}$
 is not affected by the noise variations, we obtain
 \begin{equation}
 v(t/2)[v'(t/2)]^\dag
 =\exp
 \left(-i \phi_i(t) \frac{Z_i}{2}\right)
 \exp\left(-i
\phi_j(t) \frac{Z_j}{2}
 \right)
 \end{equation}
 with
 \begin{equation}
  \phi_{k}(t)=
\int_0^{t/2}
\epsilon_{k}(t')dt'
  -
  \int_{t/2}^t\epsilon_{k}(t') dt',
 \end{equation}
and $k=i,j$.  This type of expression resembles filtered noise contributions in spin echo sequences
 ~\cite{yoneda_999-fidelity_2018}.
 We can in particular define a
spin-echo Ramsey contrast $C_\mathrm{se}(t)=\mathbb{E}[\exp
 \left(-i \phi_k(t)\right)]$, and we obtain
 the average gate channel
 \begin{equation}
  \mathcal{E}_{ZZ}(\rho)
  =
  R_{ZZ}(\theta)
\big(
  [\Phi\otimes\Phi](\rho)
\big)
R_{ZZ}^\dag(\theta)
 \end{equation}
with
 \begin{align}
    \Phi (\sigma)
    &= \frac{1+C_\mathrm{se}(t)}{2}\sigma
    +\frac{1-C_\mathrm{se}(t)}{2}Z\sigma Z
    =\big(1-p(t)\big)\sigma
    +p(t) Z\sigma Z\, ,
\end{align}
and $p(t) = \big(1-C_\mathrm{se}(t)\big)/2$. The noise part $\Phi\otimes\Phi$
of the channel corresponds in the first order in $p\ll 1$ to the superoperator
\begin{equation}
  \mathcal{S}=
  \big(1-2p(t)\big)
  \left[
    (I\otimes I)\otimes(I\otimes I)
    \right]
    +p(t)
    \left[
      (I\otimes Z) \otimes (I\otimes Z) + (Z\otimes I) \otimes (Z\otimes I)
    \right],
\end{equation}
and we obtain the process fidelity
$\mathcal{F}_\mathrm{pro}
=1-2p(t)$ and the average gate fidelity
\begin{equation}
  \mathcal{F}
  =
  \frac{4\big(1-2p(t)\big)+1}{5}=\frac{5-8p(t)}{5}=\frac{1+4C_\mathrm{se}(t)}{5}\,.
\end{equation}

 \subsubsection{Limiting cases}
 The first interesting situation is quasi-static noise, where spin-echo is known
 to exactly cancel the effect of noise~\cite{yoneda_999-fidelity_2018}.
 In our framework, this can be proven easily using the fact that
 $\epsilon_{i,j}(t')=\epsilon_{i,j}(t'+t)$, which implies $\phi_{i,j}(t)=0$, and
 thus $\mathcal{F}=1$. In other words, our implementation is robust to quasi-static
 noise.

 In the opposite limit of white noise, spin-echo is not useful
 as $\epsilon_{i,j}(t')$ and $\epsilon_{i,j}(t'+t)$ are statistically independent
 variables, and we obtain $C_\mathrm{se}(t) =\exp(-2t/T_2^*)$,
 and thus the average gate fidelity
 \begin{equation}
  \mathcal{F}=\frac{1+4\exp(-2t/T_2^*)}{5}\,.
 \end{equation}

\end{document}